\documentclass[  reprint,amsmath,amssymb,aps,]{revtex4-2}
\usepackage[unicode]{hyperref}
\usepackage{amsmath}
\usepackage[usenames]{color}
\usepackage{color}
\usepackage{colortbl}
\definecolor{linkcolor}{rgb}{0.8,0,0.2}
\definecolor{citecolor}{rgb}{0,0.6,0.2}
\definecolor{urlcolor}{rgb}{0,0,1}
\hypersetup{
    colorlinks, linkcolor={linkcolor},
    citecolor={citecolor}, urlcolor={urlcolor}}
\usepackage{graphicx}
\usepackage{epstopdf}
\usepackage[utf8]{inputenc}
\usepackage[english]{babel}

\usepackage{amsfonts}
\usepackage{eufrak}
\usepackage{mathrsfs}
\usepackage{bm}

\begin{document}


\title{Two-dimensional dynamic beam steering by Tamm plasmon polariton}
\author{Rashid G. Bikbaev$^{1,2}$}
\author{Kuo-Ping Chen$^{3,4}$}
\author{Ivan V. Timofeev$^{1,2}$}
\affiliation{$^1$Kirensky Institute of Physics, Federal Research Center KSC SB
RAS, 660036, Krasnoyarsk, Russia}
\affiliation{$^2$Siberian Federal University, Krasnoyarsk 660041, Russia}
\affiliation{$^3$College of Photonics, National Yang Ming Chiao Tung University, Tainan 711, Taiwan}
\affiliation{$^4$Institute of Photonics Technologies, National Tsing Hua University, Hsinchu 300, Taiwan}

\date{\today}

\begin{abstract}
The dynamic steering of a beam reflected from a photonic structure supporting Tamm plasmon polariton is demonstrated.
The phase and amplitude of the reflected wave are adjusted by modulating the refractive index of a transparent conductive oxide layer by applying a bias voltage.
It is shown that the proposed design allows for two-dimensional beam steering by deflecting the light beam along the polar and azimuthal angles.
\end{abstract}

\maketitle

\section{Introduction}

The control of light wave amplitude and phase is crucial to design light detection and ranging device (LIDAR). 
Significant advances in subwavelength technologies, such as photo- or electronic lithography, have made it possible to create solid-state lidars~\cite{Li2022,Kim2021,JulianoMartins2022}, in which the optical properties of light are controlled by metasurfaces - structures consisting of sub-wavelength elements. 
Metasurfaces open up new opportunities for implementation of holograms~\cite{Huang2018}, lenses~\cite{Bosch2021}, media with anomalous reflection~\cite{Sun2012,Li2015}, lasers~\cite{Sroor2020,Xu2021},  perfect absorber with critical coupling~\cite{Liang2020,Bikbaev2023}, sensors ~\cite{Wu2022,Maksimov2022} etc. 
An actively tuned metasurface with control of phase and amplitude of individual elements ensures the generation of an arbitrarily complex wave front. There are several approaches to actively rearrange optical properties of a photonic structure. 
The first one is infiltration of liquid crystals under the metasurface~\cite{Li2020,Su2017,Chen2016,Belyaev2008}. 
The advantage of this method is the control over optical properties via variation of both electric field and temperature. 
However, one of the key drawbacks of such devices is their switching time as large as several milliseconds. 
Another promising approach is application of phase-change materials, such as: silicon~\cite{Yang2023}, vanadium dioxide~\cite{Hashemi2016,Yang2022} and GST~\cite{8351784}. 
In this case switching occurs 3 orders of magnitude faster than in structures based on liquid crystals, but this switching is discrete. 
Particular attention should be paid to the investigation held by Prof. Atwater and Dr. Huang Y.-W. groups, in which the possibility of replacing conventional semiconductor materials with transparent conductive oxides was demonstrated ~\cite{Huang2016,Thureja2020}. 
It has been shown that the applied bias voltage decreases the real part of the dielectric permittivity of the conductive oxide. As a result, a significant phase jump near the gap plasmon resonant wavelength is provided. 
Similar effect was demonstrated in \cite{Bikbaev2022}, here the gap plasmon resonance was replaced by the Tamm plasmon polariton (TPP)~\cite{Kaliteevski2007,F_WU_2023_Gr_Abs,Vyunishev2017,Chen_2022_TPP_UL_APR,Chen_2023_TPP_GS_APR}. 
The attractiveness of the proposed structure consists in the fact that the use of a multilayer mirror makes it possible to excite high-quality resonances that are more sensitive to changes of structure parameters. 
In this paper 2D beam steering by the metasurface - photonic crystal based structure is presented. 
The metasurface from square nanobricks separated from the photonic crystal by graphene, sapphire and ITO films.

\section{Description of the model}
\begin{figure}[h]
   \centering
    \includegraphics[width=90mm]{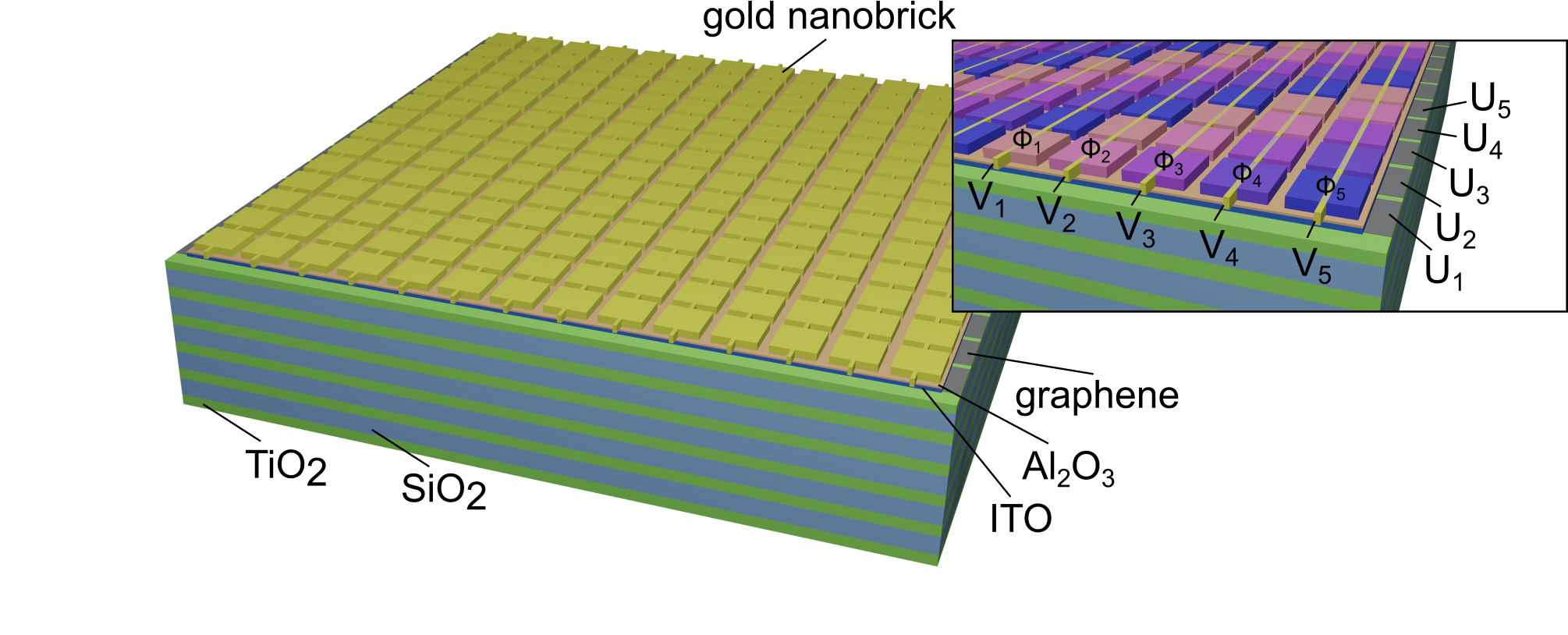}
    \caption{Tamm plasmon polariton based structure for two-dimensional dynamic beam steering. 
    } 
    \label{fig1}
\end{figure}

A schematic representation of structure under study is shown in Fig.~\ref{fig1}.

The metasurface is formed of gold~\cite{Johnson1972} square nanobricks with width $w = 190$~nm, height of $h = 95$~nm and pitch $p = 200$~nm. 
The width of the contact connecting the nanobricks is 10 nm.
The metasurface is located on sapphire layer with refractive index $n_{Al_{2}O_{3}} = 1.74$ and thickness $d_{Al_{2}O_{3}} = 5$~nm deposited on a 20 nm ITO layer.
The entire structure is placed on the surface of the photonic crystal coated with a graphene monolayer.
The unit cell of the photonic crystal is formed of titanium dioxide and silicon dioxide with refractive indexes $n_a = 2.45$, $n_b = 1.44$ and thicknesses $d_a = 135$~nm, $d_b = 165$~nm, respectively.
The number of photonic crystal periods $N_{\mathrm{PhC}} = 6$.

The distribution of the electron density in the thin ITO layer as a function of applied bias voltage has been calculated by numerical solution of Poisson and drift-diffusion equations. 
In this case the ITO layer has been presented as a semiconductor with the bandgap of $E = 2.8$~eV, electron affinity of $\chi = 5$~eV, effective electron mass of $m^* = 0.25 m_e$, permittivity $\varepsilon=9$ and initial carrier concentration of $N_{0}$ = 2.8 × 10$^{20}$ cm$^{-3}$. 
The DC permittivity of the Al$_2$O$_3$ is $\varepsilon_\mathrm{{Al_2O_3}}=9$.
The gold nanobricks and the graphene monolayer were used as electrical contacts. 
Since the ITO layer is separated from the gold contact by sapphire layer,  no current flows are observed between the contacts under bias voltage, however, this leads to a redistribution of the volume concentration of charge carriers in the ITO film. 
When there is no bias voltage between the contacts, electron depletion is observed at the ITO-Al$_2$O$_3$ boundary. 
This is due to the fact that the work function of the ITO is lower than the work function of the gold. The increase of the bias voltage between nanobricks and graphene layer leads to the displacement of charge carriers in the volume of transparent conductive oxide and their accumulation near the boundary with Al$_2$O$_3$.

\begin{figure}
    \centering
    \includegraphics[width=90mm]{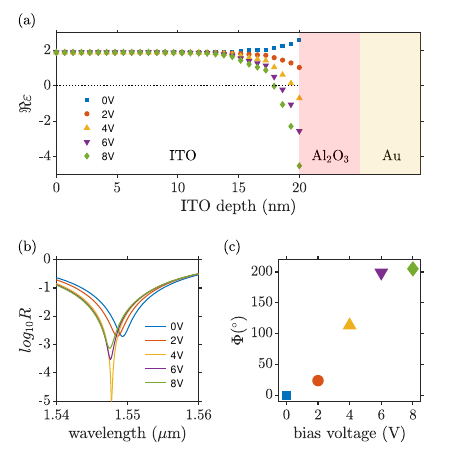}
    \caption{(a) The real part of the dielectric permittivity $\Re \varepsilon$ of the ITO layer for different applied bias voltage. (b) Reflectance spectra of the structure presented in Fig.~\ref{fig1} and  (c) simulated phase shift as a function of applied bias voltage between Au nanostripes and monolayer graphene.} 
    \label{fig2}
\end{figure}

Fig.~\ref{fig2}a shows the dependence of the real part of the complex permittivity of the ITO layer at wavelength of 1550~nm on the applied bias voltage and the distance from the Al$_2$O$_3$ boundary. 
The increase in voltage leads to a significant change in the dielectric constant of the conductive oxide in a thin 3~nm layer. 
At voltages greater than 4V, it takes values close to zero.
A further increase in voltage leads to the real part of the dielectric constant becoming negative.


Fig.~\ref{fig2}b shows the reflectance spectra of structure under study at different values of the bias voltage applied to ITO, calculated by the FDTD method. 
The minimum reflection near 1550 nm correspond to the Tamm plasmon polaritons localized at the PhC-metasurface interface. 
It can be seen that an increase in the bias voltage leads to a blue shift of the resonant wavelength. This effect can be explained by the fact that the increase in voltage leads to an increase in the volume concentration of the charge carriers near the ITO-Al$_2$O$_3$ boundary. As a result, the real part of the complex dielectric permittivity of the ITO becomes negative, and it acquires metallic properties. Thus, an additional term appears in the expression of phase matching corresponding to the phase winding when passing through the ITO layer.
\begin{equation}
    \varphi_{PhC}+\varphi_{MS}+\varphi_{ITO}=2\pi.
\end{equation}

The displacement of the TPP wavelength  into the short-wavelength region leads to a significant change in the reflection phase $\Phi$ near 1550 nm (see Fig.~\ref{fig2}c).
When the bias voltage increases from 0 to 8 volts, the phase increases by 208 degrees.
Thus, by changing the voltage applied to the ITO film, it is possible to control the phase of the wave reflected from the nanobricks. This effect can be used to control the beam in two spatial directions, since the proposed scheme allows for control of the phase of each nanobrick.

\begin{figure*}[t]
    \centering
    \includegraphics[width = 180mm]{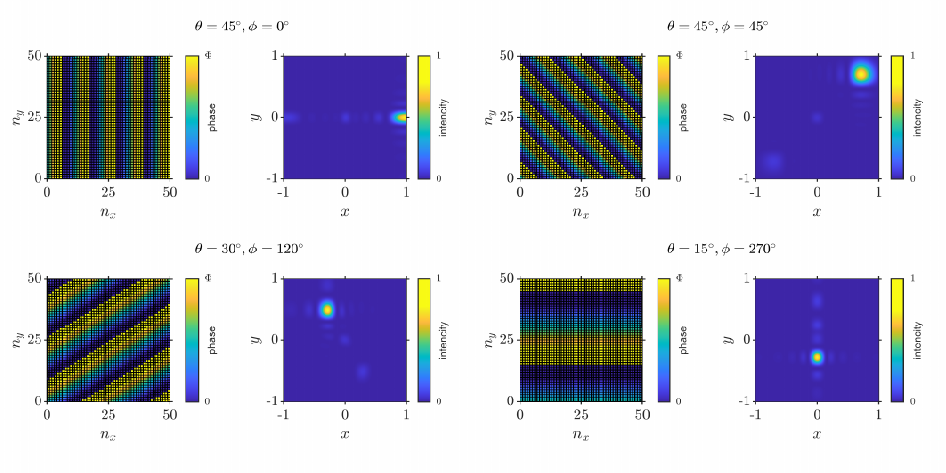}
    \caption{Phase distribution along metasurface and far field intensity for different value $\theta$ and $\phi$. In this case $x=\tan(\theta)\cos(\phi)$, $y=\tan(\theta)\sin(\phi)$, $n_x=n_y=50$ is ordinal numbers of nanobricks along $x$ and $y$-direction, respectively.} 
    \label{fig3}
\end{figure*}

\begin{figure}[t]
    \centering
    \includegraphics[width = 80mm]{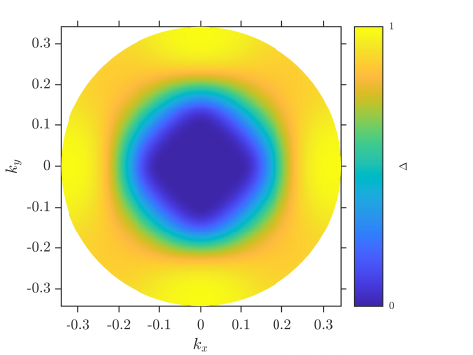}
    \caption{$\Delta = I_{N^2}-I_{2N}$ for the proposed structure.
    In this case $k_x=\cos(\phi)\sin(\theta)$, $k_y=\sin(\phi)\sin(\theta)$.} 
    \label{fig4}
\end{figure}

To demonstrate this effect, we calculated the intensity of diffraction maxima in the far field from the metasurface with $n_{x}n_{y}$ number of contacts.  
The simulation results are shown in Fig.~\ref{fig3}.
Unlike the previously proposed structure based on nanostrips~\cite{Bikbaev2022}, which provides control only by angle $\theta$, in this case it becomes possible to control the beam by azimuthal angle $\phi$.

The control of the bias voltage applied to the ITO film and the gold nanobrick allows to obtain the required phase distribution along metasurface.
Thus, it is possible to form a diffraction grating of the required period and implement intensity control of $0$ and $\pm1$ diffraction orders both in angle $\theta$ and in angle $\phi$. 
At the same time, the intensity of $+1$ order is significantly greater than the intensities of the $0$ and $-1$ orders. This is explained by the fact that the phase profiles shown in Fig.~\ref{fig3} lead to an increase in the intensity of the field only in the direction of $+1$ order, while for $0$ and $+1$ orders the waves are attenuated or quenched as a result of destructive interference.

In practice, $N^2$ number of contacts can be efficiently realized for a small number of metasurface elements~\cite{Kim2022}. 
With the control scheme shown in Fig.~\ref{fig1}, it is possible to change the reflection phase only from the $2N$ number of nanobricks. 
In this case, the necessary phase profile in the far field is provided by solving an optimization problem that allows us to properly distribute the remaining $N^2-2N$ number of phases~\cite{SabriMosallaei2022}. 
A decrease in the number of contacts imposes a restriction on polar angle $\theta$, since with its increase the intensity of the diffraction maximum decreases rapidly. 
In this regard, the coefficient $\Delta = I_{N^2}-I_{2N}$ was calculated. $\Delta$ is the difference between the intensity of the diffraction maximum in the far field provided by  $N^2$- and $2N$-contacts schemes. The calculation results are shown in Fig.~\ref{fig4}.
It can be seen that $\Delta$ is minimal for $k_x$ and $k_y$ lying in the range from -0.12 to +0.12 ($|\theta| < 6^\circ$). In this range of angles, $2N$-contacts scheme allows for the same intensity in the diffraction maximum as $N^2$. In other words, the proposed structure can be used for small-angle deflection of a light beam without loss of intensity in diffraction maxima.
\section{Conclusion}
The paper demonstrates a phase control for the wave reflected from the Tamm plasmon polariton based metasurface by modulating the refractive index in a thin layer of transparent conductive oxide located at the boundary of a one-dimensional photonic crystal and a metasurface.
It is shown that the proposed structure can be used as a dynamic phase diffraction grating. 
Calculations have shown that varying the bias voltage applied to the ITO film and gold nanobricks makes it possible to effectively control the phase of the reflected wave in two spatial directions and, as a result, deflect the incident beam both along the polar and azimuthal angles.

\section*{Funding}
This research was funded by the Russian Science Foundation (project no. 22-42-08003).
This work was supported by the National Science and Technology Council (NSTC 112-2223-E-007-007-MY3; 111-2923-E-007 -008 -MY3; 111-2628-E-007-021 ).



\bibliography{ref,references}

\end{document}